\documentclass[journal]{IEEEtran}

\usepackage{algorithm}
\usepackage{algpseudocode}
\usepackage{comment}
\usepackage{float}
\usepackage{graphicx}
\graphicspath{ {} }
\usepackage[english]{babel}
\usepackage{blindtext}
\usepackage[inline]{enumitem}
\usepackage{enumitem}
\usepackage{balance}
\usepackage{cite}
\usepackage{amsmath,amssymb,amsfonts}
\usepackage{subcaption}
\usepackage[font=footnotesize]{caption}
\usepackage{textcomp}
\usepackage{xcolor}
\usepackage{hyperref}
\usepackage{booktabs}

\def\BibTeX{{\rm B\kern-.05em{\sc i\kern-.025em b}\kern-.08em
        T\kern-.1667em\lower.7ex\hbox{E}\kern-.125emX}}
        
\begin{document}
	
\title{Large Wireless Model (LWM): \\ A Foundation Model for Wireless Channels}

\author{Sadjad Alikhani, Gouranga Charan and Ahmed Alkhateeb
\thanks{The authors are with the School of ECEE, Arizona State University. Email: \{alikhani,gcharan,alkhateeb\}@asu.edu. Part of this work has been accepted at the IEEE ICMLCN, 2025 \cite{LWM_conf2025}.}}

\maketitle

\begin{abstract}

This paper presents Large Wireless Model (LWM)---the world's first foundation model for wireless channels. Designed as a task-agnostic model, LWM generates universal, rich, contextualized channel embeddings (features) that potentially enhance performance across a wide range of downstream tasks in wireless communication and sensing systems. Towards this objective, LWM, which has a transformer-based architecture, was pre-trained in a self-supervised manner on large-scale wireless channel datasets. Our results show consistent improvements in downstream tasks when using the LWM embeddings compared to raw channel representations, especially in scenarios with high-complexity machine learning tasks and limited training datasets. This LWM's ability to learn from large-scale wireless data opens a promising direction for intelligent systems that can efficiently adapt to diverse tasks with limited data, paving the way for addressing key challenges in wireless communication and sensing systems.
\end{abstract}


\begin{figure*} [t]
    \centerline{\includegraphics[width=\textwidth]{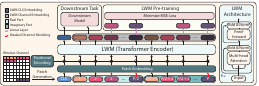}}
    \caption{This figure depicts the offline pre-training and online embedding generation process for LWM. The channel is divided into fixed-size patches, which are linearly embedded and combined with positional encodings before being passed through a Transformer encoder. During self-supervised pre-training, some embeddings are masked, and LWM leverages self-attention to extract deep features, allowing the decoder to reconstruct the masked values. For downstream tasks, the generated LWM embeddings enhance performance. The right block shows the LWM architecture, inspired by \cite{vaswani2023attentionneed}.}
    \label{fig:lwm_model}
\end{figure*}

\section{Introduction} \label{sec:intro}

Current and future wireless communication and sensing systems feature important trends that promise substantial performance gains  \cite{rappaport2019wireless,9349624}. For example, these systems are rapidly relying on the use of large antenna arrays, the operation over high frequency bands in mid-band, millimeter wave (mmWave), and sub-terahertz, the support of massive number of communicating and sensing devices of various quality of service requirements, and the densification of network infrastructure nodes. Further, these wireless communication and sensing systems increasingly interact with each other, from coordination and integration to assisting each other. Achieving the high potential of these new trends, however, requires overcoming critical challenges in high-dimensional signal processing, complex optimization problems, massive wireless overhead requirements, and intricate network management among others. All that motivates the development of novel approaches for the modeling,  optimizing, and operation of next-generation wireless communication and sensing systems. 

Traditional modeling techniques, such as statistical models and optimization-based approaches, struggle to address these challenges effectively. These methods often rely on simplified models or scenario-specific features, failing to generalize across the diverse and dynamic environments of future wireless communication and sensing networks. For instance, they may not capture complex interference patterns in dense small-cell networks or scale poorly to high-dimensional MIMO systems. Deep learning has emerged as a promising alternative, offering data-driven (model-based or model-free) solutions for optimizing network performance, resource allocation, and signal processing \cite{Shea17, Yu22, alrabeiah2020deep, alrabeiah2019deep, Jagannath21}. However, deep learning approaches also face significant limitations. First, they typically \textbf{require large labeled datasets}, which are often scarce in wireless networks and are typically expensive and hard to collect.  Second, traditional deep learning models like convolutional neural networks (CNNs) and recurrent neural networks (RNNs) struggle with specific aspects of wireless communication and sensing tasks. CNNs may not capture temporal dependencies efficiently \cite{diba2017temporal, donahue2015long}, while RNNs often struggle with long-term dependencies and real-time computational efficiency \cite{bahdanau2014neural, vaswani2023attentionneed}. These limitations underscore the need for a more robust and adaptable framework for leveraging and deploying deep learning in wireless communication and sensing networks. 

To address these challenges, we propose Large Wireless Model (LWM), a foundation model specifically designed for wireless communication and sensing channels. LWM introduces a task-agnostic framework with pre-training on large-scale synthetic data. As a task-agnostic model, \textbf{LWM serves as a universal feature extractor for multiple downstream tasks, facilitating complex problem-solving with limited labeled data}. It leverages transformer models with multi-head attention mechanisms to capture complex spatial and temporal relationships in wireless channel data. Inspired by advancements in natural language processing (NLP) \cite{vaswani2023attentionneed, devlin2019bertpretrainingdeepbidirectional, jiang2023mistral, touvron2023llama, achiam2023gpt}, audio processing \cite{baevski2020wav2vec, radford2023robust, hsu2021hubert}, and computer vision \cite{dosovitskiy2021imageworth16x16words, Khan_2022, han2022survey}, LWM learns rich, context-aware embeddings that can be utilized for various downstream wireless tasks, such as channel estimation, beamforming, and interference management. LWM is \textit{pre-trained} on extensive wireless channel datasets, covering a wide range of wireless scenarios. This approach enables the model to capture fundamental properties of wireless propagation and network dynamics, which can be transferred to real-world scenarios, even with limited task-specific data. Through these innovations, LWM addresses the key challenges of limited labeled data, complex spatial-temporal dependencies, and the need for generalization across diverse wireless environments. The key contributions can be summarized as follows:
\begin{itemize}
	\item {We introduce the world's first foundation model for wireless channel embeddings, capable of extracting universal, rich, and context-aware features from complex wireless channels and in diverse environments.}
	\item {We demonstrate the LWM's effectiveness across multiple downstream tasks, showcasing its ability to generalize to various wireless scenarios with limited task-specific data.}
	\item {We provide a comprehensive analysis of the LWM's performance compared to conventional approaches that use raw wireless channels, highlighting its advantages in feature extraction and generalization.}
	
\end{itemize}

This work introduces a new framework for leveraging and deploying deep learning in wireless communication and sensing systems by leveraging the power of foundation models to address key modeling, design, and deployment challenges in wireless systems. The pre-trained LWM model, scripts, datasets, demo, and instructions are available on the Wireless Intelligence Lab's Hugging Face page \footnote{Available on Hugging Face: \href{https://huggingface.co/wi-lab}{https://huggingface.co/wi-lab}}, allowing researchers to incorporate them into their projects.

\section{Prior Work} \label{sec:prior_work}
Foundation models have transformed artificial intelligence by introducing a paradigm of large-scale pre-training followed by task-specific fine-tuning \cite{baevski2020wav2vec, radford2023robust, hsu2021hubert, jiang2023mistral, touvron2023llama, achiam2023gpt, dosovitskiy2021imageworth16x16words}. These models, exemplified by Bidirectional Encoder Representations from Transformers (BERT) \cite{devlin2019bertpretrainingdeepbidirectional} in NLP and Wav2Vec 2.0 \cite{baevski2020wav2vec} in audio processing, leverage transformer architectures with multi-head attention mechanisms to capture complex relationships in data. The pre-training phase typically involves massive datasets and novel learning objectives: BERT uses self-supervised methods, namely masked language modeling and next-sentence prediction, while Wav2Vec 2.0 employs contrastive learning and masked prediction tasks for audio understanding. This process allows the models to learn rich, contextual representations of their input domains. The key to their success lies in the attention mechanism, which enables dynamic focus on relevant parts of the input, and the scale of pre-training, which allows the model to capture a wide range of patterns and relationships. The resulting pre-trained model serves as a powerful feature extractor, capable of being fine-tuned on various downstream tasks with limited task-specific data. This transfer learning capability, combined with the model's ability to capture long-range dependencies and generalize across scenarios, has led to state-of-the-art performance across numerous applications.

Unlike traditional models, which process sequences sequentially (like RNNs \cite{10.1162/neco.1997.9.8.1735}) or in a restricted local context (like CNNs \cite{he2015deepresiduallearningimage}), Transformers look at all parts of a sequence simultaneously. This parallel processing allows each element to relate to every other element, capturing dependencies across the entire input sequence. For example, in a sentence, words like "bank" and "river" might appear far apart, but self-attention lets the model understand their association when the context implies a natural setting, rather than financial. By calculating attention scores between each pair of words, Transformers allow each word to dynamically "attend" to others, helping the model to understand nuanced meanings and relationships.

In language models like BERT, each word token in a sentence is converted into a dense vector representation (embedding) and is associated with a query, key, and value. The attention mechanism then calculates a weighted average of these values for each word, where the weights are derived from the similarity between the queries and keys. This process allows the model to focus on important words and phrases depending on context. For instance, in the sentence, “The animal didn’t cross the street because it was too tired,” the model can determine that “it” refers to “the animal” rather than “the street.” This disambiguation is achieved because self-attention scores between “it” and “the animal” would be higher than between “it” and “the street.”

Another example can be seen in GPT models \cite{achiam2023gpt}, where the attention mechanism not only understands contextual nuances but also generates coherent text. Given a prompt, the model iteratively predicts the next word by attending to all previous words, ensuring that each generated word maintains context. For instance, when prompted with “Once upon a time in a small village,” GPT can generate a follow-up that maintains the narrative theme by giving more weight to probable story elements over unrelated concepts. This process enables the model to construct coherent, contextually relevant text, showcasing the power of the self-attention mechanism to understand complex dependencies and generate highly fluent and context-aware language output.

Moreover, in vision models, Transformers leverage the same self-attention mechanism to capture spatial relationships across pixels or image patches, providing a powerful alternative to convolutional neural networks (CNNs). For example, in Vision Transformers (ViTs) \cite{dosovitskiy2021imageworth16x16words}, an image is divided into small, fixed-size patches, each of which is flattened and linearly embedded, similar to tokens in a sentence. These embeddings are then processed in parallel, allowing the Transformer to learn relationships between distant regions of an image. This capability is particularly useful for tasks requiring an understanding of global context, such as recognizing objects within complex backgrounds or interpreting spatial patterns across an entire scene. For instance, in a complex scene containing a forest with animals partially obscured by trees, CNNs \cite{he2015deepresiduallearningimage} might struggle to understand the spatial relationship between scattered animal parts due to their limited receptive fields. However, a ViT can relate these distant regions by directly computing attention scores between patches, allowing it to “see” the entire animal even if parts of it are distant in pixel space. This method enables ViTs to recognize high-level patterns and achieve state-of-the-art performance in tasks like object detection, image segmentation, and visual question answering, where a global understanding of the image context is essential.

The success of foundation models in other domains presents compelling opportunities for \textbf{wireless communications and sensing}, particularly in addressing challenges related to complex spatial-temporal dependencies and limited labeled data. However, adapting this paradigm to wireless systems requires overcoming several hurdles. Unlike text or images, wireless signals have unique characteristics such as complex-valued data, rapid temporal variations, and domain-specific noise patterns. The lack of large-scale, diverse datasets in wireless communications comparable to those in NLP or computer vision poses another challenge. Despite these obstacles, a foundation model for wireless communications and sensing could potentially provide a \textbf{universal feature extractor} for several tasks. By pre-training on extensive datasets (either synthetic datasets generated through advanced ray-tracing simulations or real-world data), such a model could capture fundamental properties of wireless propagation and network dynamics. This approach could enable robust performance across diverse wireless environments, even with limited task-specific data, addressing key challenges in the field and potentially revolutionizing the field.

\section{LWM} \label{sec:LWM}

Building upon the foundation model paradigm discussed in Section~\ref{sec:prior_work}, LWM applies these principles to the domain of wireless communications (and sensing). LWM is designed as a task-agnostic model for generalized feature extraction in wireless channels, addressing the unique challenges of this field. LWM adopts a self-supervised Transformer architecture with multi-head attention mechanisms and is pre-trained on a large dataset of wireless channels. LWM processes input channels in patches, enabling it to serve as a universal feature extractor for various wireless communication and sensing tasks. This approach allows LWM to capture complex patterns and provide rich, contextual representations of wireless environments, potentially improving performance across diverse scenarios even with limited task-specific data. 

LWM is built upon several core design principles:

\textbf{Patch-Based Processing:} Wireless channels are segmented into patches, enabling LWM to capture both local and global patterns efficiently. This patch-based structure allows for spatial and spectral dependencies to be encoded in a way that mimics human perception of relevant wireless features, enhancing LWM's utility as a universal feature extractor.
  
\textbf{Self-Supervised Pre-Training:} Unlike traditional supervised models that require labeled data for each task, LWM is trained on a large dataset of unlabeled wireless channels using self-supervised techniques. By leveraging masked channel modeling and an attention mechanism, LWM learns to capture complex structural relationships within the data without relying on labeled datasets.

\textbf{Multi-Head Attention:} The attention mechanism in LWM allows it to selectively focus on relevant patterns in wireless channels, dynamically assigning importance to different parts of the input. This is particularly beneficial for wireless applications, where meaningful features can vary spatially and spectrally across environments.

\textbf{Task Flexibility and Transferability:} The representations produced by LWM are highly versatile, making it possible to apply the model to numerous downstream tasks with minimal or no fine-tuning. As a result, LWM can generalize across different scenarios and geographic regions, achieving robust performance even when task-specific data is sparse or highly variable.

In the following sections, we detail each component of LWM’s pipeline, from data preprocessing and embedding to the architecture of the Transformer encoder blocks. These sections will demonstrate how LWM processes raw wireless channels into enriched feature representations and embeddings, as illustrated in Fig.~\ref{fig:lwm_model}, and provide insight into how this model bridges the gap between traditional wireless communication techniques and advanced, general-purpose AI architectures.

\section{Data Preprocessing: Masking and Embedding \label{subsec:preprocess}}  
Given the input wireless channels, we first preprocess them to align with the Transformer input format and facilitate our self-supervised pre-training method. The steps for this are outlined below.

\subsection{Patch Generation}
Unlike RNNs \cite{10.1162/neco.1997.9.8.1735}, Transformers require the entire input simultaneously. To achieve this, we feed each channel in a patch-based format \cite{dosovitskiy2021imageworth16x16words}. Each channel matrix \(\mathbf{H} \in \mathbb{C}^{M \times N}\) is split into \(P\) patches by first separating the real and imaginary components, flattening them, and then dividing each part into patches. The process is as follows:

\begin{enumerate}
\item \textbf{Separate Real and Imaginary Components:}  
   We start by separating \(\mathbf{H}\) into its real and imaginary components
   \begin{subequations}
    \begin{align}
   & \mathbf{H}_{\mathsf{real}} = \Re(\mathbf{H}), \\
   & \mathbf{H}_{\mathsf{imag}} = \Im(\mathbf{H}),
    \end{align}   
    \end{subequations}
   where \(\mathbf{H}_{\mathsf{real}} , \mathbf{H}_{\mathsf{imag}} \in \mathbb{R}^{M \times N}\).

\item \textbf{Flatten Each Component:}
   Flatten both \(\mathbf{H}_{\mathsf{real}} \) and \(\mathbf{H}_{\mathsf{imag}}\) into vectors using the vectorization operator, as follows
   \begin{subequations}
    \begin{align}
   & \mathbf{h}_{\mathsf{real}}  = \text{vec}\left(\mathbf{H}_{\text{real}}^{\top} \right), \\
   & \mathbf{h}_{\mathsf{imag}} = \text{vec}\left(\mathbf{H}_{\text{imag}}^{\top} \right),
    \end{align}   
\end{subequations}

   resulting in \(\mathbf{h}_{\mathsf{real}}, \mathbf{h}_{\mathsf{imag}} \in \mathbb{R}^{MN}\).

\item \textbf{Divide into Patches:}
   Divide each flattened component into \(P/2\) patches. Let each patch have a length \(L = 2MN/P\). We generate patches as follows
\begin{subequations}
   \begin{align}
        & \mathbf{p}_i = \mathbf{h}_{\mathsf{real}}[(i-1)L+1:iL], \quad i \in [P/2], \\
        & \mathbf{p}_{i+P/2} = \mathbf{h}_{\mathsf{imag}}[(i-1)L+1:iL], \quad i \in [P/2],
   \end{align}
\end{subequations}

where each patch \(\mathbf{p}_i \in \mathbb{R}^L\), with \(i \in [P] = \{1, 2, \dots, P\}\). 
\end{enumerate}

This patch-based approach accelerates computations, enables the model to learn both inter- and intra-patch relationships, mimics convolutional layers with self-attention, and increases design flexibility. For instance, with \(M = 32\), \(N = 32\), \(P = 128\), and \(L = 16\), this configuration generates \(128\) patches of length \(16\). Smaller patches capture fine-grained details but require higher computational cost due to the increased number of patches, and they may limit the model’s ability to learn broader structural dependencies \cite{chen2021crossvitcrossattentionmultiscalevision}. Larger patches emphasize long-range dependencies and structural patterns, reducing computational requirements but potentially overlooking finer details. Ultimately, selecting the optimal patch size depends on the task's need for local versus global information and the available computational resources, rather than any specific threshold. 

For LWM, which is designed to serve as a universal feature extractor, achieving this balance is essential. By selecting a patch size that allows both detail and structure to be preserved, LWM remains adaptable, capturing features that are sufficiently detailed to represent nuanced variations while broad enough to generalize across tasks and datasets. A useful approach for assessing an optimal patch size is to evaluate the performance of the masking strategy described next.

\subsection{Masked Channel Modeling} 
For LWM to be task-agnostic, we pre-train it in a self-supervised manner. Self-supervised learning enables LWM to capture the intrinsic structure of the data, allowing it to serve as a universal feature extractor adaptable to various downstream tasks, without reliance on labeled data. To enable this self-supervised pre-training, we propose a technique called Masked Channel Modeling (MCM).

In MCM, we mask \(p\%\) of the real part patches. This percentage is chosen to balance the model’s access to context with the challenge of reconstructing missing patches. Masking too many patches could lead to excessive information loss, making reconstruction difficult, while masking too few would provide limited incentive for the model to learn complex dependencies. Within the selected \(p\%\) of patches, specific sub-percentages are assigned:

\begin{itemize}
\item \textbf{\(80\%\) are fully masked} with a uniform vector \(\mathbf{m} = [m, m, \dots, m]^\top \in \mathbb{R}^L\), preventing the model from accessing information in those locations. This requires LWM to leverage surrounding patches to predict the masked values, pushing it to learn the spatial dependencies within the channel.
\item \textbf{\(10\%\) are replaced with random vectors} sampled from a distribution (e.g., \(\mathcal{N}(0, \sigma^2)\)), adding noise and encouraging LWM to differentiate genuine channel structures from anomalies.
\item \textbf{\(10\%\) are left unchanged}, providing partial ground truth to stabilize the model’s predictions and helping it recognize real patterns among masked and altered patches.
\end{itemize}

This masking strategy is carefully applied to both the real and imaginary parts to prevent information leakage between them. Specifically, the imaginary patches selected for masking are the exact counterparts of the randomly selected real patches. This ensures that if a real part patch is masked, the corresponding imaginary part patch is also masked, preventing any indirect inference. If only the real part of a patch \(\mathbf{p}_i\) were masked while leaving its imaginary part \(\mathbf{p}_{i+P/2}\) unmasked, the model could leverage the unmasked imaginary patch \(\mathbf{p}_{i+P/2}\) to predict \(\mathbf{p}_i\). By masking both components of each selected patch, LWM learns each component’s structure independently, resulting in robust feature extraction and reducing the chance of unintended data leakage.

The selected \(p\%\) of patches are masked before going through the input embedding, positional encoding, and finally the LWM pre-training stages. The corresponding high-dimensional embeddings of masked patches at the output of LWM then pass through a simple linear layer that maps each embedding back to the original patch size. The goal is to minimize the Mean Squared Error (MSE) between the reconstructed masked patches and their original values, expressed as

\begin{equation} 
\mathcal{L}_{\mathsf{MCM}} = \frac{1}{|\mathcal{M}|} \sum_{i \in \mathcal{M}} \left\| \mathbf{W}^\mathsf{dec} \mathbf{e}^\text{LWM}_i - \mathbf{p}_i \right\|^2, \label{eq:loss}
\end{equation}
where
\(\mathcal{M}\) represents the set of all selected (masked) patches,
\(\mathbf{e}^\mathsf{LWM}_i \in \mathbb{R}^{D}\) is the high-dimensional embedding of the \(i\)-th masked patch at LWM's output,
\(\mathbf{W}^\mathsf{dec} \in \mathbb{R}^{L \times D}\) is the weight matrix of the linear layer used to map \(\mathbf{e}_i\) back to the original patch size,
and \(\mathbf{p}_i\) is the original value of the \(i\)-th patch.

This approach allows LWM to develop highly refined embeddings that can be decoded accurately using a simple linear layer, underscoring the richness and expressiveness of the learned representations. Because the Transformer encoder does not know which patches will be masked or replaced by random patches, it is forced to maintain a contextual representation for every input patch, ensuring the embeddings are robust and contextually aware. Additionally, as random replacement only occurs for a small fraction of all patches (\(10\%\) of the \(p\%\) masked patches), the model’s ability to capture spatial and structural dependencies remains unaffected \cite{devlin2019bertpretrainingdeepbidirectional}. The high performance of these linear layers in reconstructing masked patches highlights the quality of LWM’s output embeddings, which effectively capture complex spatial relationships within the channel, enabling efficient decoding and ensuring robust performance across a variety of downstream tasks.

\subsection{CLS Patch} We prepend an additional patch, known as the CLS (classification) patch \cite{devlin2019bertpretrainingdeepbidirectional}, to the sequence of channel patches, increasing the sequence length to \(P+1\). The CLS patch is initialized as a learnable vector, denoted \(\mathbf{p}_{\text{CLS}} = [c, c, \dots, c]^\top \in \mathbb{R}^L\), where \(c\) is set as a random value. Through its interactions across the Transformer layers, it aggregates and summarizes information from all other patches in the sequence.

This interaction enables the CLS patch to capture a comprehensive view of the entire input by attending to each patch in each layer, accumulating information from both local details and broader structures. As the sequence passes through multiple Transformer layers, the CLS patch’s representation is refined, with each layer integrating lower-level details (e.g., spatial or temporal variations) and higher-level features (e.g., overall channel quality or dominant paths). The resulting representation, \(\mathbf{e}_{\mathsf{CLS}}^{\mathsf{LWM}}\), serves as a compact, high-level summary of the input sequence
\begin{equation}
\mathbf{e}_{\mathsf{CLS}}^{\mathsf{LWM}} = f_\mathsf{LWM}(\mathbf{e}_{\mathsf{CLS}}, \{\mathbf{e}_i\}_{i=1}^{P}),
\end{equation}
where \(f_\mathsf{LWM}\) represents the transformations and interactions within LWM. Through this mechanism, the CLS patch identifies the patches that contribute most significantly to the structure and dependencies within the channel, highlighting the segments most critical in defining the channel’s overall characteristics.

Additionally, the compact size of the CLS patch makes it an efficient, low-dimensional encoded representation of the channels. This compactness is advantageous, as it serves as an expressive summary without requiring further model training, given that LWM has already been pre-trained in a task-agnostic manner to capture these global features.

The \textbf{attention scores} assigned between the CLS patch and each patch \(\mathbf{p}_i\) reveal the relative importance of each patch, making the CLS patch a valuable interpretability tool across various wireless tasks. For example, in \textbf{channel estimation}, the CLS patch can highlight critical segments of the channel, such as those representing major multipath components, essential for accurate channel estimation. In \textbf{classification tasks} such as line-of-sight (LoS) and non-line-of-sight (NLoS) classification, the CLS patch effectively summarizes the input, capturing the distinguishing features needed to identify LoS versus NLoS conditions. By attending to patches with high relevance to signal paths, reflection, and scattering, the CLS patch representation \(\mathbf{e}_{\text{CLS}}^{\text{LWM}}\) captures a robust global context that is ideal for classification tasks, where understanding the overall channel state is crucial. Additionally, in \textbf{resource allocation}, the CLS patch’s attention scores \(\{\alpha_{\text{CLS},i}\}_{i=1}^{P}\) (where \(\alpha_{\text{CLS},i}\) is the attention weight of the \(i\)-th patch with respect to the CLS patch) help identify channel segments that demand more resources. This optimization supports spectrum and power allocation by prioritizing the most influential patches. In \textbf{beamforming} and \textbf{beam selection}, the CLS patch can focus on patches that carry directional cues, assisting in adaptive beam selection by highlighting segments most indicative of optimal beam configurations. In \textbf{semantic communications}, the CLS patch enables efficient encoding by capturing patches containing the most contextually relevant information, enhancing communication quality while reducing redundancy. These examples demonstrate the versatility of the \textbf{CLS patch} across wireless applications, where it serves as an adaptable \textbf{focal point} for understanding and prioritizing channel features that directly impact performance \cite{wu2023clstokenneedzeroshot}.

\subsection{Input Embedding} 
After masking and prepending the CLS patch to the channel patch sequence, each patch is projected into an embedding space with dimension \( D \) using a linear layer, effectively mapping each flattened patch to a \( D \)-dimensional vector. Given a set of patches \(\{\mathbf{p}_\text{CLS}, \mathbf{p}_1^\mathsf{m}, \mathbf{p}_2^\mathsf{m}, \dots, \mathbf{p}_{P}^\mathsf{m}\}\), where \(\mathbf{p}_i^\mathsf{m} \in \mathbb{R}^{L}\) represents a masked patch (with only \(15\%\) of patches masked), the linear layer performs the following transformation for each patch \( \mathbf{p}_i^m \)
\begin{equation}
\mathbf{e}_i^\mathsf{emb} = \mathbf{W}^\mathsf{emb} \mathbf{p}_i^\mathsf{m} + \mathbf{b} \in \mathbb{R}^{D}, \quad i \in \{\mathsf{CLS}\} \cup [P],
\end{equation}
where \(\mathbf{W}^\mathsf{emb} \in \mathbb{R}^{D \times L}\) is the weight matrix, and \(\mathbf{b} \in \mathbb{R}^{D}\) is the bias vector for all patches. This transformation produces the initial patch embeddings \(\mathbf{E}^\mathsf{emb} = [\mathbf{e}_\mathsf{CLS}^\mathsf{emb}, \mathbf{e}_1^\mathsf{emb}, \mathbf{e}_2^\mathsf{emb}, \dots, \mathbf{e}_{P}^\mathsf{emb}]^\top \in \mathbb{R}^{(P+1) \times D}\), allowing the Transformer to process all patches in a common high-dimensional feature space where relationships can be effectively captured.

Embedding patches into a higher-dimensional space before feeding them into the Transformer is essential for capturing complex relationships, especially compared to models like autoencoders, which often reduce dimensionality. A higher-dimensional embedding provides a richer, more expressive feature space, enabling each patch to retain fine-grained information about the channel data. 
Directly using the original patch size as an embedding would limit the \textbf{model’s capacity} to capture meaningful relationships. The choice of a higher embedding dimension \( D \) enables the Transformer to establish detailed contextual relationships, similar to how embeddings in text-based models capture the semantic relationships between words. In language models, embeddings allow each token (word or subword) to represent not only its isolated meaning but its meaning contextualized by surrounding words. This context-awareness is essential for handling nuances like polysemy. 

Likewise, in wireless channels, high-dimensional embeddings allow each patch to capture its information with respect to the entire channel, representing dependencies between different parts of the channel and capturing structural nuances.
In wireless channels, fine-grained details such as spatial correlations, multipath effects, and scattering are crucial for accurate representation. By embedding patches into a higher dimension, we ensure sufficient capacity to capture these intricate dependencies. Lower-dimensional representations, as in autoencoders, often sacrifice such details for compression, which can limit the Transformer's capacity to understand the channel’s complex structure.

Furthermore, \textbf{positional encodings} are added to these embeddings to provide the Transformer with information about the order of patches, as it inherently lacks sequence awareness. To achieve this, we define a positional encoding patch \(\mathbf{p}_i^{\mathsf{pos}} \in \mathbb{R}^{L}\) for each patch \(i\) as follows:

\begin{itemize}
\item For the CLS token, the positional encoding patch is a uniform vector of zeros.
\item For each subsequent patch \(i \in [P]\), the positional encoding patch is a uniform vector filled with the value \(i\), providing an incremental encoding. Mathematically, this is expressed as

\begin{equation}
\mathbf{p}_i^{\mathsf{pos}} = i \cdot \mathbf{1}_{L}, \quad i \in [P],
\end{equation}

where \(\mathbf{1}_{L}\) is a vector of ones in \(\mathbb{R}^{L}\). 
\end{itemize}
Each positional encoding patch \(\mathbf{p}_i^{\mathsf{pos}}\) is then mapped into the embedding space using a learned embedding matrix \(\mathbf{W}^{\mathsf{pos}} \in \mathbb{R}^{D \times L}\) and a bias vector \(\mathbf{b}^\mathsf{pos} \in \mathbb{R}^{D}\), resulting in positional encodings

\begin{equation}
\mathbf{e}_i^{\mathsf{pos}} = \mathbf{W}^{\mathsf{pos}} \mathbf{p}_i^{\mathsf{pos}} + \mathbf{b}^\mathsf{pos}, \quad i \in \{\mathsf{CLS}\} \cup [P].
\end{equation}
These position embeddings are then added to the corresponding patch embeddings \(\mathbf{e}_i\), yielding the position-encoded input embeddings

\begin{equation}
\mathbf{e}_i^{\mathsf{input}} = \mathbf{e}_i^\mathsf{emb} + \mathbf{e}_i^{\mathsf{pos}}, \quad i \in \{\mathsf{CLS}\} \cup [P].
\end{equation}
This addition forms the final set of position-encoded embeddings \(\mathbf{E}^{\text{input}} = [\mathbf{e}_\text{CLS}^{\mathsf{input}}, \mathbf{e}_1^{\mathsf{input}}, \mathbf{e}_2^{\mathsf{input}}, \dots, \mathbf{e}_{P}^{\mathsf{input}}]^\top \in \mathbb{R}^{(P+1) \times D}\), effectively incorporating ordered context into each patch embedding for the Transformer model.

\section{Model Architecture}
We adopt the Transformer encoder architecture from \cite{vaswani2023attentionneed}, with modifications tailored to the context of wireless channels. The model processes input embeddings through a sequence of \( E \) encoder blocks, progressively refining the embeddings to capture increasingly complex relationships and contextual patterns in the data. Across the \( n \)-th encoder block, where \( n \in [E] \), the input embeddings \(\mathbf{E}^{\mathsf{input}}_n \in \mathbb{R}^{(P+1) \times D}\) evolve, with each layer enhancing feature richness. This sequence culminates in the final output embeddings, denoted as \(\mathbf{E}^\mathsf{LWM} \in \mathbb{R}^{(P+1) \times D}\), which constitute the LWM embeddings, encapsulating a refined representation suitable for downstream tasks. 

Within each encoder block, the input embeddings go through the following components sequentially: \textbf{multi-head attention}, \textbf{layer normalization} with \textbf{residual connections}, a \textbf{feed-forward network}, and another layer normalization with residual connections. Each component refines the embeddings by capturing different aspects of the data structure. The output from one encoder block serves as the input to the next block, enabling the model to progressively build complex representations that capture the dependencies within the wireless channels. The details of these model components \textit{at each encoder block} are outlined below.

\subsection{Self-Attention Mechanism}  

The self-attention mechanism allows each patch in the input sequence to assign contextually weighted importance to all other patches, achieved through a dot-product similarity measure. Given input embeddings \( \mathbf{E}^{\mathsf{input}} \in \mathbb{R}^{(P+1) \times D} \), where \(P\) is the number of patches and \(D\) is the embedding dimension, each row of \(\mathbf{E}^{\mathsf{input}}\) represents an embedding vector corresponding to a patch.

To compute self-attention, we first derive the \textit{Query} (\(\mathbf{Q}\)), \textit{Key} (\(\mathbf{K}\)), and \textit{Value} (\(\mathbf{V}\)) matrices using linear transformations, each defined by learned weights
\begin{equation}
\mathbf{Q} = \mathbf{E}^{\mathsf{input}} \mathbf{W}^Q, \quad \mathbf{K} = \mathbf{E}^{\mathsf{input}} \mathbf{W}^K, \quad \mathbf{V} = \mathbf{E}^{\mathsf{input}} \mathbf{W}^V,
\end{equation}
where \(\mathbf{W}^Q, \mathbf{W}^K, \mathbf{W}^V \in \mathbb{R}^{D \times D^\prime}\). Here, \(D^\prime\) is typically set to \(D\) for single-head attention but may vary to \(D_\mathsf{H}\) for multi-head attention, as will be discussed next.

The core of self-attention begins by calculating the \textbf{scaled dot-product} of the query (\(\mathbf{Q}\)) and key (\(\mathbf{K}\)) matrices, capturing the similarity among patches in the input sequence. This similarity measure is computed as follows

\begin{equation}
\mathbf{S} = \frac{\mathbf{Q} \mathbf{K}^\top}{\sqrt{D^\prime}},
\end{equation}
where \(\mathbf{S}\) represents the scaled similarity scores, and the scaling factor \(\sqrt{D^\prime}\) prevents gradient issues like explosion or vanishing. These scaled similarity scores are then normalized by applying the softmax function

\begin{equation}
\mathbf{A} = \text{softmax}(\mathbf{S}),
\end{equation}
where \(\mathbf{A}\) is the attention weight matrix, which ensures that the weights assigned to each patch sum to 1 for each query, allowing the model to focus selectively on the most relevant patches. Finally, the attention weights are used to compute a weighted sum of the value (\(\mathbf{V}\)) matrix, producing the attention output

\begin{equation}
\text{Attention}(\mathbf{Q}, \mathbf{K}, \mathbf{V}) = \mathbf{A} \mathbf{V}.
\end{equation}

This process enables the model to dynamically adjust its focus based on the contextual relevance of each patch in relation to others, effectively capturing both local and global relationships across the input sequence.

The Query, Key, and Value matrices in the self-attention mechanism represent distinct, interrelated aspects of each input patch, working together to capture both local and global dependencies in the data. Each of these matrices has a specific purpose, ultimately contributing to how the model emphasizes certain relationships over others within the input sequence.

\textbf{Query (\(\mathbf{Q}\)):} The Query matrix encodes the \textbf{intent} or \textbf{interest} of each patch in the sequence, representing the way each patch seeks information relevant to itself from other patches. Queries help determine what aspects of the surrounding data are most pertinent to the current patch, as each query vector in \(\mathbf{Q}\) essentially acts as a \textbf{question} that asks how similar or relevant other patches are to it.

\textbf{Key (\(\mathbf{K}\))}: The Key matrix complements the Query matrix by encoding \textbf{characteristics} of each patch that can be evaluated by queries from other patches. While Queries seek information, Keys provide the attributes or features by which each patch can be "queried." For any two patches, their query and key vectors are compared to assess their mutual relevance, which is captured through the dot product \( \mathbf{Q} \mathbf{K}^\top \). High similarity scores indicate that certain patches carry information of high interest to one another.

\textbf{Value (\(\mathbf{V}\)):} The Value matrix represents the actual information that each patch provides to the model, contributing to the weighted sum in the final self-attention calculation. Once the model identifies relevant patches (through the similarity scores from the dot product of Queries and Keys), the corresponding Value vectors are aggregated based on their importance, defined by the computed attention weights. Thus, Values are akin to the \textbf{content} that is passed along in the attention mechanism, shaping the contextualized representation of each patch by integrating relevant details from others.

To better understand the roles of Query, Key, and Value, consider the analogy of a conference setting. Imagine each participant has specific interests or questions (Queries) they want addressed, such as AI techniques or wireless advancements. Each participant also has unique expertise (Keys) they can offer. If someone’s interest in AI aligns with another’s data science expertise, a strong match occurs, enabling an exchange of knowledge (Values). In self-attention, this relationship is formalized by computing the similarity between each Query and all Keys. When a Query and Key pair strongly align, a higher attention weight is assigned, resulting in a weighted aggregation of the corresponding Value vectors. This effectively allows each patch to gather information most relevant to its context.

Together, Queries, Keys, and Values facilitate a process where each patch determines which other patches to \textbf{attend} to, assigns importance to each based on their contextual relevance, and then updates its representation based on a weighted sum of the relevant information. This arrangement allows the Transformer model to dynamically focus on relationships across patches, enabling it to capture both immediate and extended dependencies in the data with great flexibility and precision.

\subsection{Multi-Head Attention}
\begin{figure}[t]
    \centering
    \includegraphics[width=\columnwidth]{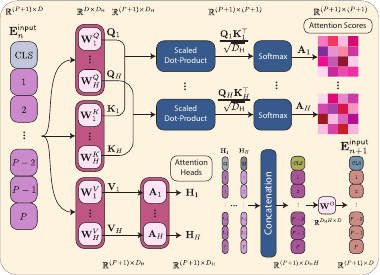}
    \caption{Multi-head Attention Mechanism}
    \label{fig:attention}
\end{figure}

Multi-head attention builds upon self-attention by enabling the model to learn multiple representations simultaneously, as depicted in Fig. \ref{fig:attention}. Instead of a single attention head, \(H\) independent self-attention mechanisms are applied, each with distinct learned weight matrices
\begin{equation}
\text{head}_h = \text{Attention}(\mathbf{E}^{\mathsf{input}} \mathbf{W}_h^Q, \mathbf{E}^{\mathsf{input}}\mathbf{W}_h^K, \mathbf{E}^{\mathsf{input}}\mathbf{W}_h^V),
\end{equation}
where \(\mathbf{W}_h^Q, \mathbf{W}_h^K, \mathbf{W}_h^V \in \mathbb{R}^{D \times D_\mathsf{H}}\) are the transformation matrices specific to each head, and \(D_\mathsf{H} = \lfloor D/H \rfloor\). By processing across multiple heads, the model learns representations from various subspaces of the data, enhancing its ability to capture nuanced relationships. The outputs of all heads are concatenated and passed through a linear transformation
\begin{equation}
\text{MultiHead}(\mathbf{Q}, \mathbf{K}, \mathbf{V}) = \text{Concat}(\text{head}_{1}, \dots, \text{head}_{H}) \mathbf{W}^O,
\end{equation}
where \(\mathbf{W}^O \in \mathbb{R}^{D_\mathsf{H} H \times D}\). This combined output from multiple heads enables the model to focus on diverse aspects of the input sequence simultaneously, enriching the feature representation by capturing different patches and interactions in parallel.

In other words, each attention head can be thought of as a unique \textbf{lens} through which the model interprets the input data. In a multi-faceted problem space, these different lenses enable the model to pick up on varying levels of detail and relational structures simultaneously. To extend our analogy of participants in a conference: imagine that each participant not only has a set of questions (Queries) and expertise (Keys) but can also approach the conversation from multiple viewpoints or specializations (e.g., technical details, future trends, industry applications). One head might focus on broad concepts (like general channel structure or coarse spatial relationships), while another might pick up fine-grained, detailed aspects (such as precise timing or frequency patterns). 

\subsection{Feed-Forward Network (FFN)} 
Each encoder layer has a feed-forward network (FFN) that processes individual patch embeddings independently. Formally, the FFN applies two linear transformations separated by a non-linear ReLU activation
\begin{equation}
\text{FFN}(\mathbf{e}_i) = \max(\mathbf{0}, \mathbf{e}_i \mathbf{W}_{1} + \mathbf{b}_{1}) \mathbf{W}_{2} + \mathbf{b}_{2},
\end{equation}
where \(\mathbf{W}_{1} \in \mathbb{R}^{D \times D_{\mathsf{FF}}}\) and \(\mathbf{W}_{2} \in \mathbb{R}^{D_{\mathsf{FF}} \times D}\). The intermediate layer size \(D_{\mathsf{FF}} = T D\) expands the embedding dimension by a factor of \(T\), which empirically improves the model’s expressiveness. The FFN enriches each embedding with more complex transformations, and because each embedding is processed separately, the FFN layer does not introduce dependencies across patches.

\subsection{Layer Normalization and Residual Connections}
Layer normalization stabilizes the training by ensuring that each layer’s output has zero mean and unit variance
\begin{equation}
\text{LayerNorm}(\mathbf{e}_i) = \frac{\mathbf{e}_i - \mu_i}{\sigma_i},
\end{equation}
where \(\mu_i\) and \(\sigma_i\) denote the mean and standard deviation across the feature dimension of each patch embedding \(\mathbf{e}\). This normalization is essential for convergence, especially in deeper networks, as it reduces covariate shifts.

Residual connections are added around each sub-layer (self-attention and FFN) to improve gradient flow
\begin{equation}
\text{Output}_i = \text{LayerNorm}(\mathbf{e}_i + \text{sub-layer}(\mathbf{e}_i)).
\end{equation}
The addition operation allows gradient information to flow more directly through the network, mitigating vanishing gradient problems and enabling efficient backpropagation.

\section{Pre-Training}
Given the data preprocessing steps and the LWM model architecture described in the previous two subsections, we pre-train this model to be then leveraged in multiple downstream tasks. The following subsections outline the key components of LWM’s pre-training process.

\subsection{Pre-Training Dataset}
The LWM pre-training process utilizes a vast and diverse dataset of over 1 million wireless channels from 15 scenarios within the DeepMIMO dataset \cite{Alkhateeb2019}, with $80\%$ of the data dedicated to training and $20\%$ to validation. These scenarios include \textit{O1}, \textit{Boston5G}, \textit{ASU Campus}, \textit{New York}, \textit{Los Angeles}, \textit{Chicago}, \textit{Houston}, \textit{Phoenix}, \textit{Philadelphia}, \textit{Miami}, \textit{Dallas}, \textit{San Francisco}, \textit{Austin}, \textit{Columbus}, and \textit{Seattle}. The first three scenarios are significantly larger, while the remaining twelve city scenarios average around $1500$ effective users each. In wireless communications and sensing, relationships within the data often manifest as dependencies across various dimensions, requiring the model to capture both local patterns and broader spatial structures. Transformers are most effective when trained on large-scale and diverse datasets \cite{dosovitskiy2021imageworth16x16words}, as these enable them to generalize and learn complex relationships. Training on this extensive dataset ensures that LWM captures nuanced wireless patterns, making it highly capable for a variety of downstream tasks.

\subsection{Wireless Channel Adaptation for Transformer Architecture} 
The wireless channels in the adopted dataset are structured as \((M, N) = (32, 32)\) matrices, where \(N\) corresponds to the number of subcarriers and \(M\) to the number of antennas. To prepare this data for Transformer processing, each channel matrix \(\mathbf{H} \in \mathbb{C}^{32 \times 32}\) is split into \(P = 128\) patches: \(64\) from the real part and \(64\) from the imaginary part, as detailed in section (\ref{subsec:preprocess}). Each patch contains values from \(L = 16\) consecutive subcarriers out of the 32 total subcarriers, spanning each antenna, resulting in patches of size \(16 \times 1\).

For self-supervised learning, our proposed Masked Channel Modeling (MCM) technique is applied as described in section (\ref{subsec:preprocess}). Specifically, \(9\) out of \(64\) real-part patches are randomly masked (approximately \(15\%\) of patches), with the corresponding imaginary-part patches masked similarly to prevent data leakage. Masking is conducted based on three probability-based actions: with probability \(0.1\), patches are replaced with a random vector; with probability \(0.8\), they are replaced with a uniform MASK array (value \(m = 1\)); and with probability \(0.1\), they remain unchanged. This approach enables LWM to learn robust contextual relationships across both masked and unmasked patches, facilitating effective feature extraction from wireless channels. Importantly, because LWM does not know which patches are masked, it is compelled to understand all inter- and intra-patch relationships, rather than focusing solely on the masked patches. Although we only optimize the loss function based on the masked patches, this structure ensures that LWM captures a comprehensive representation across the entire input, reinforcing its ability to generalize effectively. The objective is to minimize the prediction loss (\ref{eq:loss}) by selecting LWM embeddings \(\mathbf{e}^\mathsf{LWM}_i\) of the selected (masked) patches \(\mathcal{M}\) and minimizing

\begin{equation}
\min_{\mathbf{W}^\mathsf{dec}, \mathbf{\Theta}}  \sum_{i \in \mathcal{M}} \left\| \mathbf{W}^\mathsf{dec} \mathbf{e}^\mathsf{LWM}_i - \mathbf{p}_i \right\|^2,
\end{equation}
where \(\mathbf{\Theta}\) represents the parameters of the LWM model. This optimization encourages the model to learn mappings that accurately reconstruct masked patches from the embeddings, refining its ability to capture critical context from both masked and unmasked channel sections.

After masking, a CLS patch vector \(\mathbf{p}_\mathsf{CLS}\), initialized with random values, is prepended to the start of the channel patch sequence. Each \(16 \times 1\) patch vector is then projected into a \(D = 64\)-dimensional space via a linear layer, yielding embeddings \(\mathbf{e}_i^{\mathsf{emb}} \in \mathbb{R}^{64}\) for each patch. These embeddings are further enriched with positional encodings to capture structural dependencies critical for learning within wireless channel data, as discussed in section (\ref{subsec:preprocess}). The final input embedding matrix \(\mathbf{E}^\mathsf{input} \in \mathbb{R}^{129 \times 64}\) consolidates spatial and frequency information from both real and imaginary components, ready for processing through LWM's transformer-only encoder.

\subsection{Pre-Training Loss Function}

LWM uses \textbf{MSE loss} instead of the \textbf{cross-entropy loss} commonly found in NLP models like BERT. In language models, cross-entropy loss is effective for predicting discrete tokens from a fixed vocabulary, as in Masked Language Modeling (MLM), where masked tokens are classified based on surrounding context
\begin{equation}
\mathcal{L}_{\mathsf{MLM}} = - \sum_{i \in \mathcal{M}} \log p(y_i | \mathbf{e}_{i}^{\mathsf{BERT}}),
\end{equation}
where \(\mathcal{M}\) represents masked positions, \(y_i\) is the true token, and \(p(y_i | \mathbf{e}_{i}^{\mathsf{BERT}})\) is the predicted probability. This approach works for discrete data but is unsuitable for continuous-valued wireless channels, which lack a fixed vocabulary. Instead, LWM treats masked channel modeling (MCM) as a \textbf{regression task}, using MSE loss (\ref{eq:loss}) to measure the error between predicted and actual masked patch values. This allows LWM to learn spatial and temporal dependencies from surrounding unmasked patches, developing robust feature representations tailored to the continuous nature of wireless channels.

\begin{table}
    \begin{center}
    \caption{LWM Pre-training Setup Parameters}
    \label{tab:lwm_parameters}
    \begin{tabular}{| c | c |}
    \hline
    \textbf{Parameter} & \textbf{Value} \\
    \hline
    Antennas at BS (\(N\)) & \(32\) \\
    \hline
    Antennas at UEs & \(1\) \\
    \hline
    Subcarriers (\(M\)) & \(32\) \\
    \hline
    Patch Size (\(L\)) & \(16\) \\
    \hline
    Embedding Size (\(D\)) & \(64\) \\
    \hline
    Channel Patches (\(P\)) & \(128\) \\
    \hline
    Attention Heads (\(H\)) & \(12\) \\
    \hline
    Encoder Layers (\(E\)) & \(12\) \\
    \hline
    FFN Hidden Size (\(D_\mathsf{FF}\)) & \(256\) \\
    \hline
    Head Dimension (\(D_\mathsf{H}\)) & \(5\) \\
    \hline
    Masking Percentage (\(p\)) & \(15\) (\(80/10/10\)) \\
    \hline
    Learning Rate & \(1 \times 10^{-4}\) \\
    \hline
    Batch Size & \(64\) \\
    \hline
    Optimizer & Adam \\
    \hline
    Adam \(\beta_1\) & \(0.9\) \\
    \hline
    Adam \(\beta_2\) & \(0.999\) \\
    \hline
    Adam \(\epsilon\) & \(1 \times 10^{-8}\) \\
    \hline
    Weight Decay & \(1 \times 10^{-5}\) \\
    \hline
    Dropout Rate & \(0.1\) \\
    \hline
    Model Parameters & \(600\text{K}\) \\
    \hline
    Training Set Size & \(820\text{K}\) \\
    \hline
    Validation Set Size & \(200\text{K}\) \\
    \hline
    \end{tabular}
    \end{center}
    \end{table}

\subsection{Pre-Training Setup Parameters}
As shown in table \ref{tab:lwm_parameters}, the pre-training setup for LWM includes \(12\) attention heads, \(12\) encoder layers, an embedding size of \(64\), and an FFN hidden size of \(256\). Training begins with a learning rate of \(1 \times 10^{-4}\), decreasing by \(10\%\) every \(10\) epochs to ensure smooth convergence. A batch size of 64 is utilized, along with the Adam optimizer (\(\beta_1 = 0.9\), \(\beta_2 = 0.999\), \(\text{eps} = 1 \times 10^{-8}\)), and a weight decay of \(1 \times 10^{-5}\) to reduce overfitting. The \(12\)-head attention mechanism enables the model to capture multiple relationships within the data, while the depth of the encoder layers allows it to extract both local and global patterns in the channels. This setup ensures effective learning, balancing convergence and generalization across various wireless communication scenarios.

By the end of pre-training, LWM is capable of producing rich, contextual embeddings from raw wireless channels. The integration of channel preprocessing, self-supervised learning, bidirectional attention, and multi-head attention mechanisms enables the model to generalize effectively across a range of scenarios, making it a powerful feature extractor for diverse downstream tasks in wireless communications and sensing systems.

\section{Inference}

LWM excels in generating rich, context-aware embeddings from raw wireless channels in real-time, with no need for additional training for embedding generation. Pre-trained on large, diverse datasets using a self-supervised, Transformer-based approach, it allows users to immediately obtain high-quality, low- and high-dimensional embeddings suitable for a wide array of downstream tasks. The pre-trained model can be employed as-is, leveraging these embeddings directly, or choose to fine-tune the model’s last layers to extract highly task-specific, fine-grained features. This flexibility makes LWM highly adaptable, enabling it to capture both local and global patterns and perform effectively across general and specialized scenarios, even in data-scarce environments.

The inference process begins with segmenting the raw wireless channel data into patches, embedding them, and adding positional encodings, similar to the pre-training phase but \textbf{without the need for masking or weight updates}. This structured approach allows LWM to extract and represent multi-scale patterns from both small and large contexts within the data. The resulting embeddings, \(
\mathbf{E}^{\text{LWM}} = [\mathbf{e}_\text{CLS}^{\mathsf{LWM}}, \mathbf{e}_1^{\mathsf{LWM}}, \mathbf{e}_2^{\mathsf{LWM}}, \dots, \mathbf{e}_{P}^{\mathsf{LWM}}]^\top = \begin{bmatrix} \mathbf{C} & \mathbf{E}^\mathsf{T} \end{bmatrix}^\mathsf{T} \in \mathbb{R}^{(P+1) \times D}\), consist of the \textbf{CLS embedding} \(\mathbf{C} \in \mathbb{R}^{D}\) and the \textbf{channel embeddings} \(\mathbf{E} \in \mathbb{R}^{P \times D}\), as shown in Fig. \ref{fig:lwm_model}.

The CLS embedding provides a holistic channel representation, making it ideal for general tasks like LoS/NLoS classification, while channel embeddings, four times larger than the input, capture intricate spatial and frequency dependencies for more detailed applications. This inference setup offers key advantages, including immediate usability, as pre-trained embeddings can be directly applied without retraining, enabling rapid deployment in resource-limited scenarios. LWM also captures both local and global channel patterns, ensuring adaptability to tasks requiring fine-grained or high-level insights. Additionally, its flexibility allows users to either apply embeddings as-is or fine-tune the last layers for task-specific improvements without retraining the full model. Furthermore, LWM generalizes well even in data-scarce environments, leveraging its pre-trained representations to maintain high performance with minimal labeled data. By balancing detailed feature extraction with computational efficiency, LWM provides a practical and adaptable solution for various wireless communication tasks.

\section{Downstream Task Evaluation}  
In this section, we show that LWM CLS and channel embeddings outperform raw wireless channels across various downstream tasks. We evaluate LoS/NLoS classification and sub-6 GHz to mmWave beam prediction as examples, though many other tasks can be explored.

\subsection{Sub-\(6\) to mmWave Beam Prediction}  
This task aims to predict the strongest mmWave beam at the receiver from a predefined codebook at the base station, based on Sub-\(6\) GHz channels. This is a specific case of \textit{channel mapping}, where instead of directly predicting mmWave channels, the model learns the relationship between Sub-\(6\) GHz channels and the best mmWave beam. Ground-truth optimal beams are computed for each user, generating a labeled dataset. The task complexity is varied by adjusting the codebook size, ranging from \(16\) to \(256\) beams. Performance is evaluated across different training data percentages to assess how well the model generalizes with less data. This makes the task highly practical for modern communication systems, as it reduces the overhead of full mmWave channel estimation. It also tests how well LWM embeddings capture the spatial and propagation characteristics of Sub-\(6\) GHz channels and generalize to higher-frequency mmWave beams, making it an ideal benchmark for comparing the efficiency and generalizability of LWM embeddings against raw channels.

\textbf{Downstream Model}: To ensure fair comparison, we use a similar downstream model complexity for both embeddings and raw channels, selecting a model optimized for raw channels as the performance benchmark. A uniform Residual $1$D-CNN architecture with $500\text{K}$ parameters is employed, designed to capture complex patterns through residual connections and weight-sharing while avoiding overfitting. The model consists of an initial convolution layer followed by three residual blocks, each containing several convolution layers that extract deeper features. This is followed by global average pooling and fully connected layers for final classification. When parameters exceed $500\text{K}$, the model overfits to raw channels, which is why we consider this architecture the benchmark for raw channels.

\textbf{Downstream Task Dataset:} The test set comprises six new scenarios from the DeepMIMO dataset, which were not part of LWM's pre-training: \textit{Denver}, \textit{Fort Worth}, \textit{Oklahoma}, \textit{Indianapolis}, \textit{Santa Clara}, and \textit{San Diego}, comprising a total of \(14840\) samples. For each user, the corresponding $28$GHz channels are generated, and the optimal beams are computed to create a labeled dataset, where $3.5$GHz channels serve as inputs and $28$GHz beams as labels. The dataset is split into \(70\%\) for training, \(20\%\) for validation, and \(10\%\) for testing. For LWM embeddings, the raw channels are first to generate LWM embeddings, the raw channels are processed through the pre-trained LWM model in real-time, with the channel embedding component $\mathbf{E}$ serving as the input for this downstream model, as shown in Fig. \ref{fig:lwm_model}.

\begin{figure}[t] 
    \centering
    \begin{subfigure}[b]{0.24\textwidth}  
        \centering
        \includegraphics[width=\textwidth]{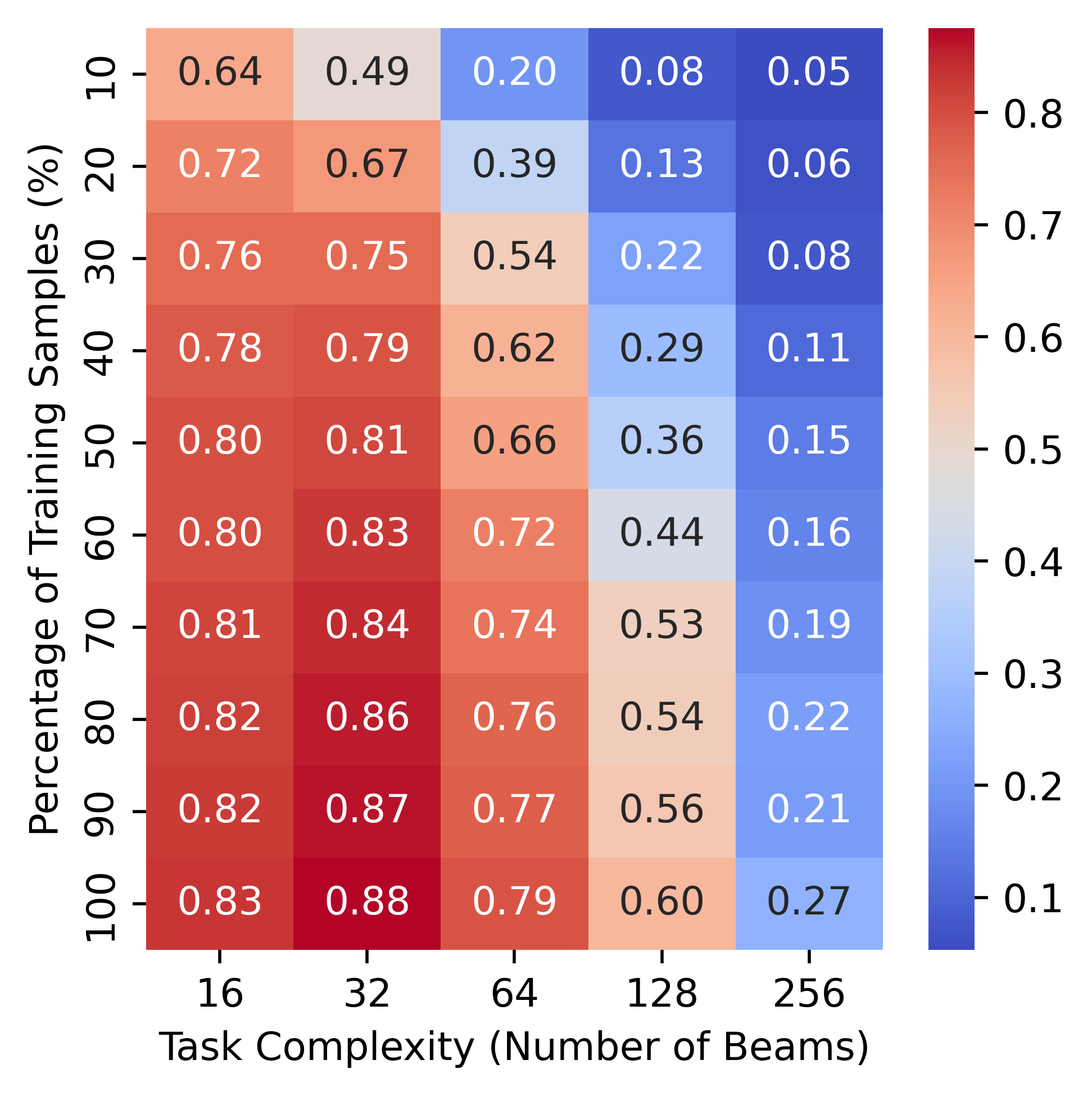}
        \caption{Raw Channels Performance}
        \label{fig:eval_1}
    \end{subfigure}
    \hfill
    \begin{subfigure}[b]{0.24\textwidth}
        \centering
        \includegraphics[width=\textwidth]{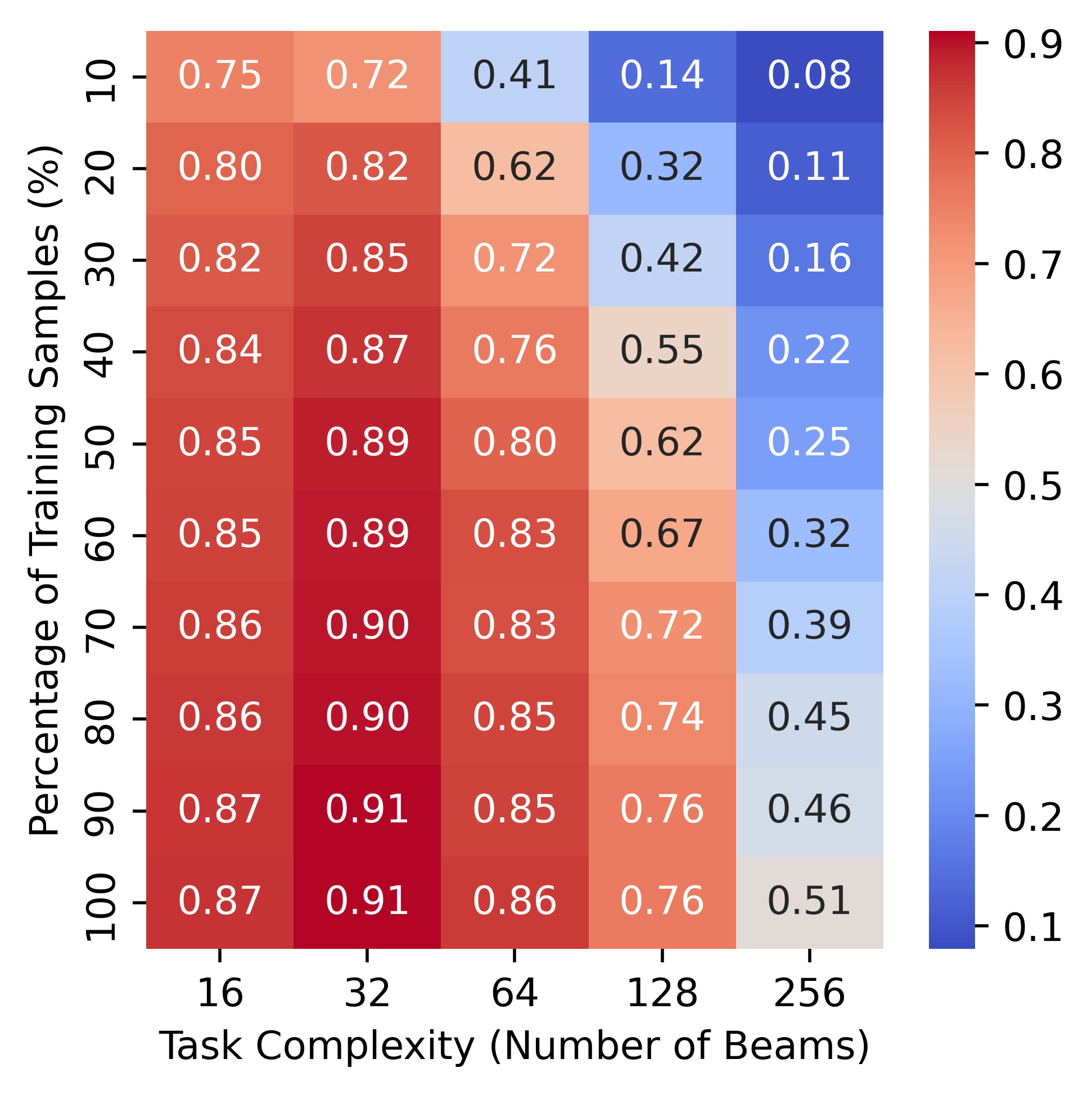}
        \caption{LWM Embeddings Performance}
        \label{fig:eval_2}
    \end{subfigure}
    \hfill
    \begin{subfigure}[b]{0.24\textwidth}
        \centering
        \includegraphics[width=\textwidth]{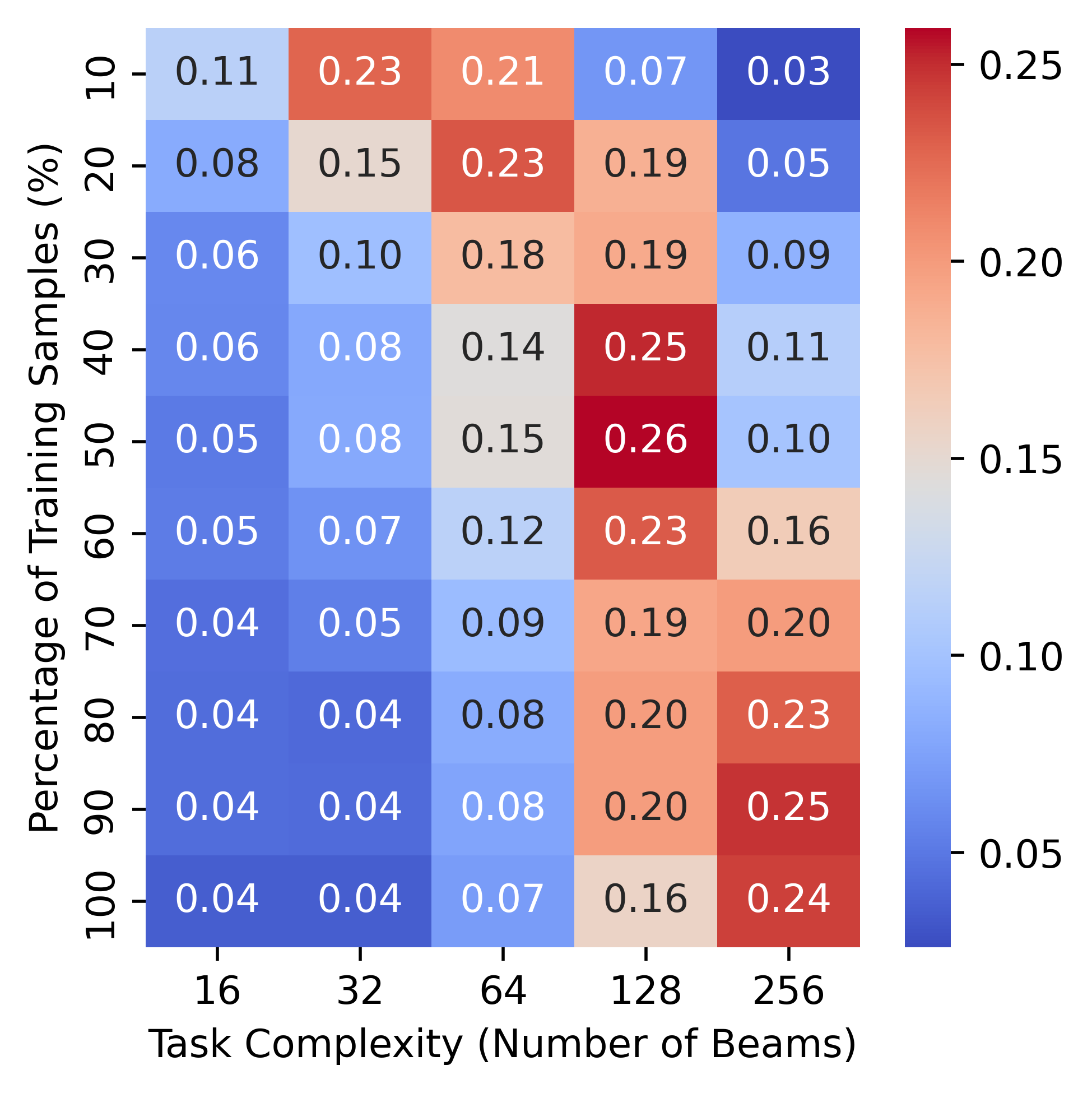}
        \caption{Performance Difference}
        \label{fig:eval_3}
    \end{subfigure}
    \hfill
    \begin{subfigure}[b]{0.24\textwidth}
        \centering
        \includegraphics[width=\textwidth]{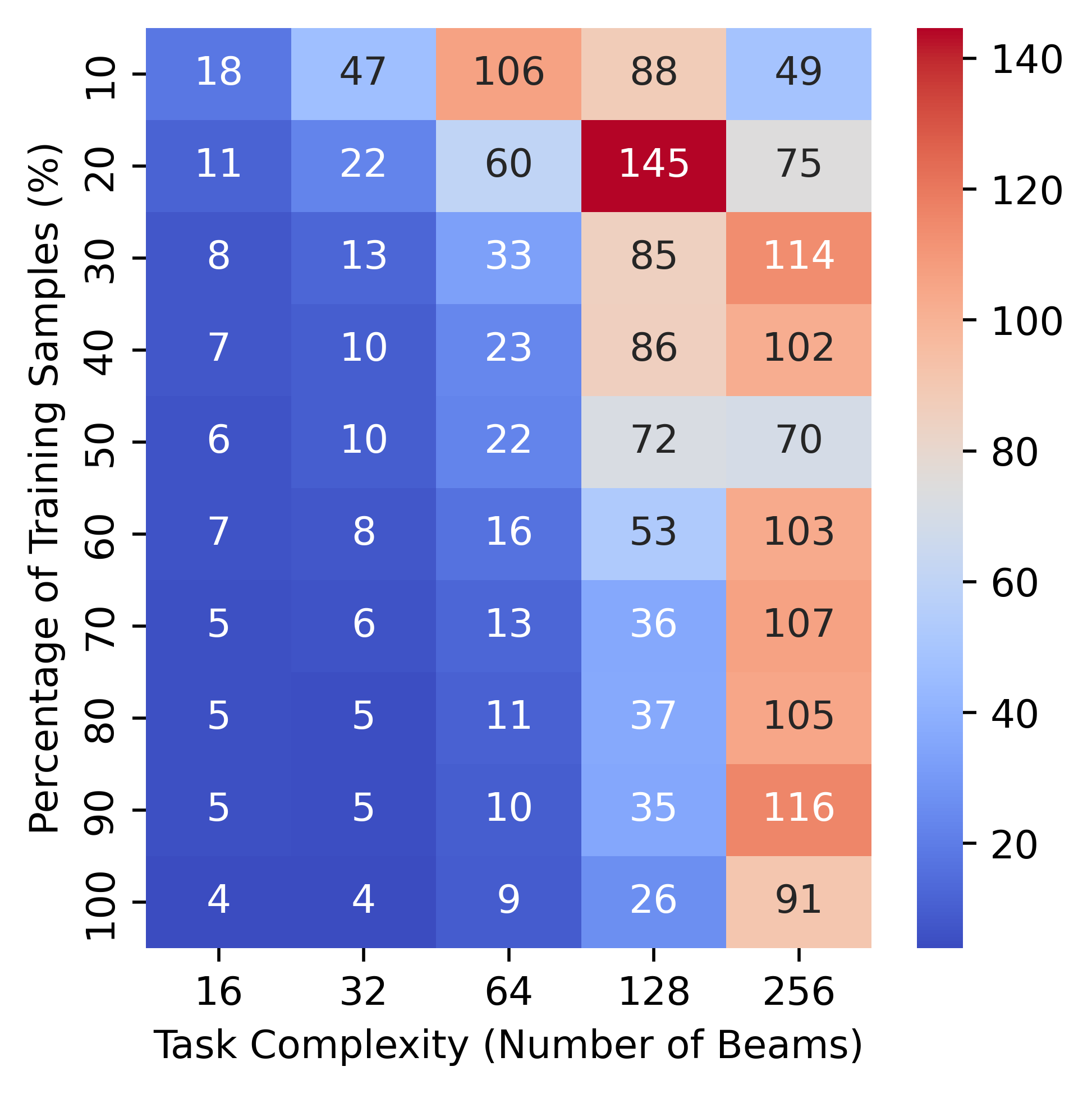}
        \caption{Percentage of Performance Gain}
        \label{fig:eval_4}
    \end{subfigure}
    
    \caption{This figure compares beam prediction F1-score performance between raw channels and their inferred LWM embeddings, based on a total of 10388 training raw channels, and highlights their relative effectiveness.}
    \label{fig:eval}
\end{figure} 

\textbf{Downstream Tasks Evaluation:} 
Fig. \ref{fig:eval} presents a comparison of the performance between raw channels and LWM channel embeddings in training a model for beam prediction, evaluated across different codebook sizes and varying amounts of training data. As shown in Fig. \ref{fig:eval_1}, the downstream models trained with raw channels require significantly more data to reach high performance levels, while LWM channel embeddings usually achieve performance saturation with just \(50\%\)-\(70\%\) of the available data and consistently outperform raw channels. Notably, LWM embeddings reach the benchmark performance of raw channels with only \(40\%\)-\(50\%\) of the data, regardless of task complexity, highlighting the data efficiency of LWM embeddings.
\par We use the F1-score for classification tasks as it accounts for imbalanced labels and provides a clearer evaluation of model performance than accuracy. The F1-score difference heatmap in Fig. \ref{fig:eval_3} highlights where embeddings surpass raw channels, aiding in model optimization for complex or low-data tasks. Meanwhile, the F1-score gain percentage heatmap in Fig. \ref{fig:eval_4} emphasizes the efficiency of embeddings, showing how they scale with fewer resources. Fig. \ref{fig:eval_3} demonstrates that embeddings are most effective when data is limited relative to task complexity. The performance difference follows a concave trend, where an initial increase in data improves performance, but beyond a certain point, the difference diminishes and gradually decreases, though it never turns negative. 

As the task complexity increases, such as with a higher number of labels, larger datasets are required to fully showcase the advantages of embeddings over raw channels, allowing embeddings to reveal their full potential in capturing intricate patterns and dependencies.

\subsection{LoS/NLoS Classification}  

This task serves as a great benchmark to evaluate CLS embeddings, which provide a highly compressed but more informative representation of raw channels. These embeddings effectively capture the critical features of a channel, making them ideal for various tasks that require a holistic understanding of the channel's behavior. For instance, in applications like CSI feedback, the CLS embeddings can reduce the overhead of sending full channel data to the base station, particularly in multi-vendor environments. Instead of relying on infeasible joint autoencoder training between user equipment (UE) and base station (BS), the LWM model functions like an encoder, delivering a compact yet illustrative version of the channel. This makes it highly suitable for tasks that need a comprehensive understanding of channel characteristics while significantly reducing transmission complexity.

\textbf{Downstream Model:} The classification model serves as the downstream network for raw channels (\(2048\) features), CLS embeddings (\(64\) features), and channel embeddings (\(8192\) features), maintaining a consistent architecture across different input representations. It begins with a \(512\)-unit fully connected layer, followed by batch normalization, ReLU activation, and dropout (\(0.1\)) to enhance stability and prevent overfitting. The pattern continues through \(256\) and \(128\)-unit layers, refining the features hierarchically. The final linear layer maps the processed \(128\)-dimensional representation to LoS/NLoS classes. Batch normalization accelerates convergence, while dropout improves generalization. 

\begin{figure}[t]
    \centering
    \includegraphics[width=\columnwidth]{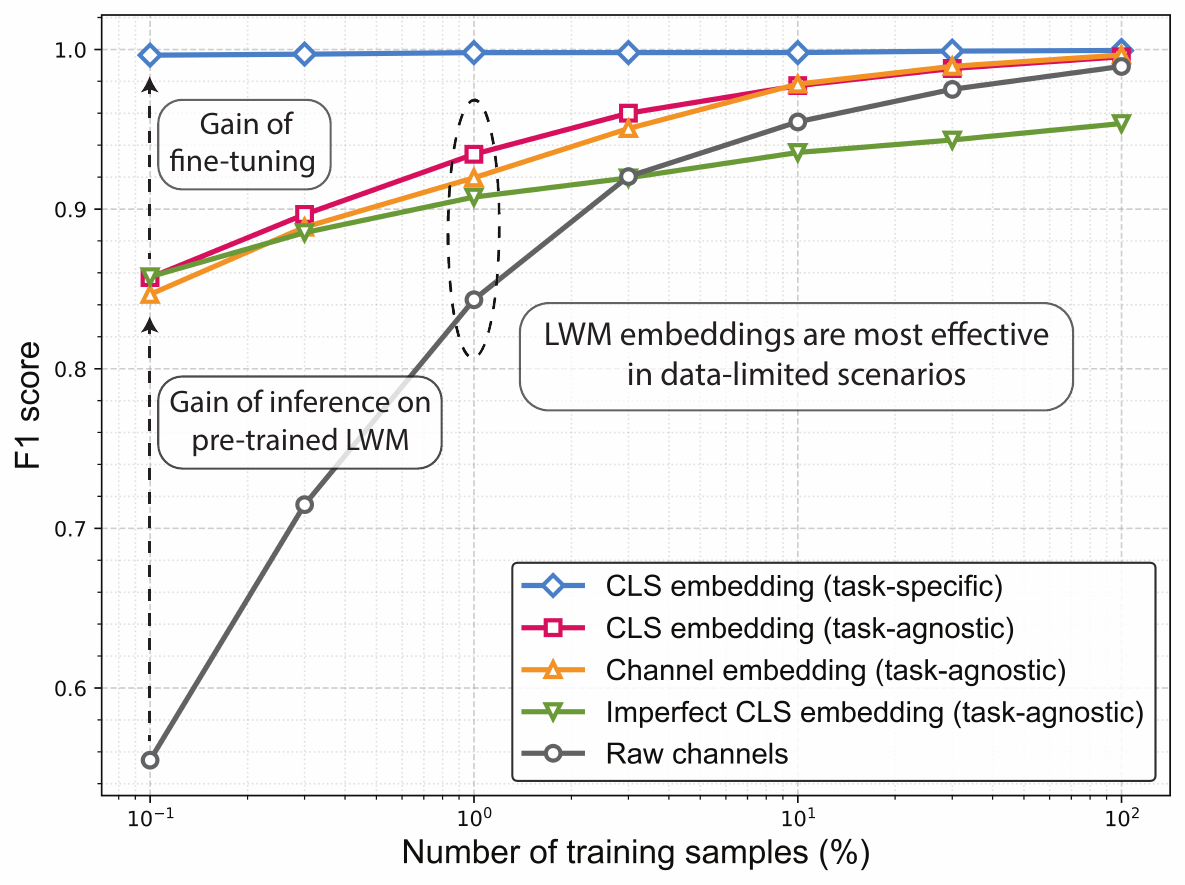}
    \caption{This figure compares F1-scores for LoS/NLoS classification using models trained on raw wireless channels and LWM embeddings across different percentages of the \(6639\) training samples. CLS embeddings are \(32 \times\) smaller than raw channels, while channel embeddings are \(4 \times\) larger. CLS embeddings inferred from noisy (imperfect) raw channels are also included to demonstrate LWM’s robustness to noise.}
    \label{fig:los}
\end{figure}

\begin{figure*} [t]
    \centerline{\includegraphics[width=\textwidth]{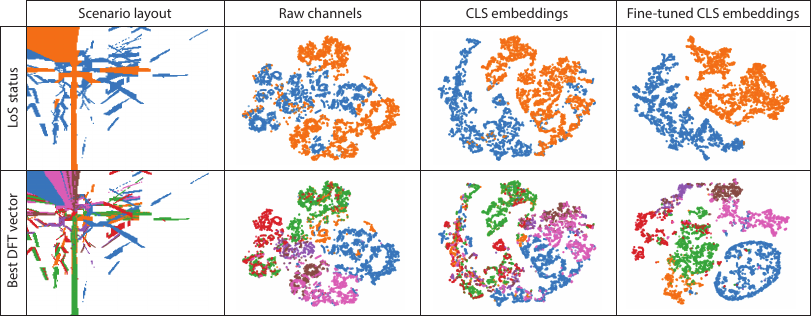}}
    \caption{This figure visualizes the distribution of users in the DeepMIMO Denver scenario based on their LoS/NLoS status (top row) and strongest DFT beam index among \(8\) beams (bottom row). User channels are projected into \(2\)D using t-SNE, comparing raw channels, task-agnostic (general-purpose) LWM embeddings, and fine-tuned LWM embeddings for each task. As seen in Fig. \ref{fig:los}, CLS embeddings clearly separate LoS and NLoS channels, enabling high zero-shot classification and strong initialization for downstream training with minimal data. Fine-tuning further enhances downstream task performance.}
    \label{fig:emb_space}
\end{figure*}

\textbf{Downstream Task Dataset:} For this task, we utilized the densified Denver dataset, where we re-ran ray tracing on the existing DeepMIMO scenario, reducing the user grid spacing from \(2.5\)m to \(1\)m. This higher density resulted in a total of \(8299\) samples, offering a more detailed representation of the wireless environment for downstream evaluations. The dataset is partitioned with up to \(80\%\) allocated for training (\(6639\) samples) and the remaining \(20\%\) for validation.

\textbf{Downstream Task Evaluation:} Fig.~\ref{fig:los} underscores LWM’s capabilities in \textit{data efficiency}, \textit{lightweight adaptation}, and \textit{noise robustness}. The downstream model is evaluated using five input types: (i) CLS embeddings (first-patch representation from frozen pre-trained LWM), (ii) channel embeddings (remaining patches of general-purpose embeddings), (iii) raw channels (unprocessed channels), (iv) CLS embeddings inferred from imperfect raw channels corrupted with complex Gaussian noise (\(\text{SNR} = 5\) dB), and (v) fine-tuned CLS embeddings. Here, \textit{general-purpose embeddings} are extracted from the pre-trained LWM without weight updates, while \textit{fine-tuned embeddings} are generated by updating only the last three layers of LWM jointly with the downstream task. This setup isolates the role of pre-trained feature hierarchies versus task-adaptive refinement. The comparison across input types quantifies LWM’s ability to balance noise suppression, data efficiency, and task-specific discriminability—key strengths for deployment in resource-constrained wireless environments.  

\begin{figure}[t]
    \centering
    \includegraphics[width=\columnwidth]{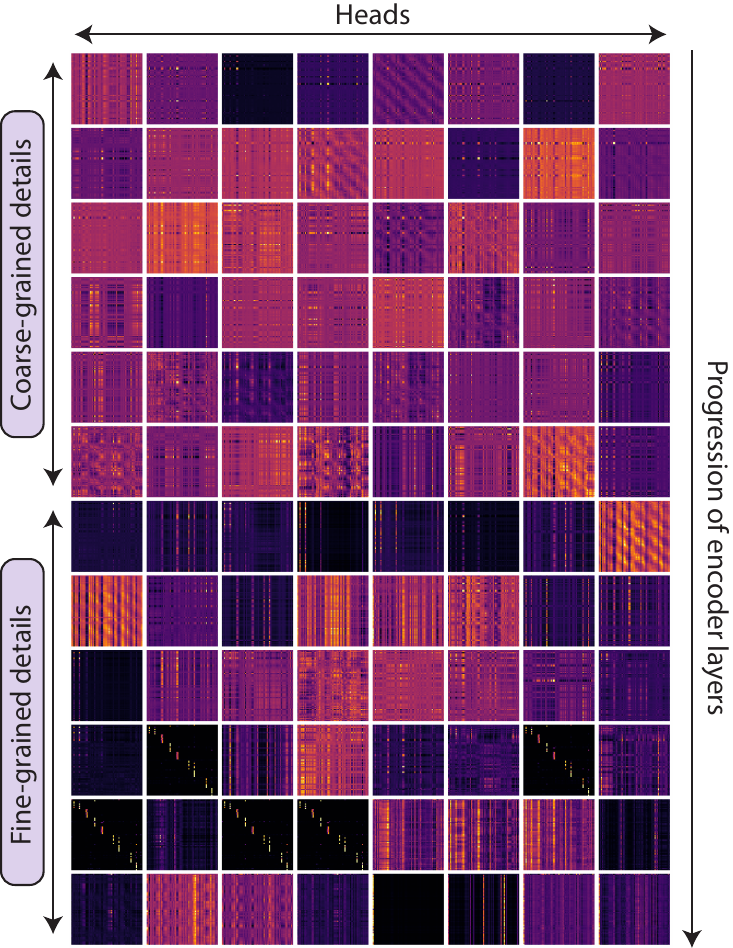}
    \caption{This figure shows attention maps of patches within a channel across LWM layers and heads. Rows represent layers, columns represent heads, and axes of each attention map denote query and key patch indices. Lighter colors indicate higher self-attention, while darker maps in lower layers suggest a focus on fine details, making them ideal for task-specific fine-tuning.}
    \label{fig:attn}
\end{figure}

The evaluation reveals several key findings. \textbf{First}, with only \textit{6 training samples}, models trained on raw channel data perform slightly better than random guessing (average F1-score = 0.55), whereas general-purpose embeddings improve performance by \textbf{+0.31 F1}, demonstrating strong class separation in the embedding space (Fig.~\ref{fig:emb_space}). \textbf{Second}, fine-tuned embeddings enable perfect class differentiation (F1 $\approx$ 1.0) even with minimal data, achieved by unfreezing and updating only the \textbf{last three layers} of LWM alongside the downstream model---a strategy grounded in empirical evidence that earlier layers encode \textbf{coarse-grained patterns} (e.g., syntax and morphology in LLMs), while deeper layers refine \textbf{task-specific details} (e.g., signal variations) \cite{tenney2019learncontextprobingsentence} (Fig.~\ref{fig:attn}). \textbf{Third}, CLS embeddings outperform channel embeddings in \textit{low-data regimes}, as channel embeddings capture complex patterns requiring larger datasets for effective utilization; however, channel embeddings slightly surpass CLS performance as training samples grow, leveraging their richer information density. \textbf{Fourth}, CLS embeddings exhibit robustness to noise, aligning closely with their noisy counterparts, which highlights the LWM’s ability to filter noise via self-attention in unseen environments. \textbf{Finally}, the fine-tuning strategy---limited to deeper layers---preserves pre-trained knowledge of coarse signal characteristics (e.g., propagation geometry), preventing overfitting on small LoS/NLoS datasets, while enabling adaptation to subtle discriminative features. For more complex tasks, such as beam prediction, full-model fine-tuning becomes necessary.  

\section{Discussion and Remarks}  

In this section, we provide additional insights related to the LWM framework, covering points on attention mechanisms, masked channel modeling based pre-training, and a comparison with autoencoder embeddings. These discussions offer complementary perspectives to enhance understanding of the model's design and functionality.

\subsection{Discussion on Attention}  
\label{sec:appendixA} 

LWM’s Transformer-based architecture leverages self-attention to model spatial and spectral dependencies in wireless channels, offering significant advantages over conventional signal processing methods. Unlike CNNs, which rely on local receptive fields, self-attention enables global feature extraction, capturing critical relationships across frequency and spatial domains in a single layer. This mechanism dynamically assigns importance scores to different patches, allowing the model to prioritize dominant channel components while suppressing noise and interference, making it well-suited for complex and dynamic wireless environments. Additionally, LWM’s bidirectional attention enables each patch to attend to both preceding and following patches, capturing inter-subcarrier dependencies and spatial correlations for improved modeling of multipath propagation.

A key strength of self-attention in LWM is its potential for \textbf{interpretability and sensitivity analysis} in wireless communications. Unlike traditional deep learning models, where feature extraction remains opaque, attention scores offer direct insight into how different channel components contribute to predictions. By analyzing these scores, wireless engineers can assess which subcarriers, antennas, or spatial regions are most influential, enabling improved resource allocation, interference mitigation, and adaptive beamforming strategies. This is particularly valuable in CSI compression, where prioritizing highly informative channel components can significantly reduce feedback overhead without sacrificing performance.  

Furthermore, multi-head attention enhances interpretability by providing a layered perspective on wireless channels. Each attention head captures different propagation characteristics—one may focus on local variations in signal strength, while another identifies global trends across subcarriers. This allows for detailed sensitivity analysis, helping researchers understand how different signal components impact model predictions under varying SNR conditions. By tracking attention shifts across different environments, engineers can evaluate robustness to channel variations, identify learning weaknesses, and refine architectures for real-world deployment.  

Overall, LWM’s self-attention framework transforms wireless feature extraction into an interpretable and adaptive process. With its ability to highlight critical channel features, facilitate sensitivity analysis, and dynamically adjust to varying conditions, LWM is not just a powerful inference model but also a valuable tool for optimizing wireless system design and performance.

\subsection{Discussion on Masked Channel Modeling} 

MCM pre-trains LWM by randomly masking subsets of input channel patches and optimizing the model to reconstruct them via contextual dependencies learned through self-attention. Unlike conventional denoising techniques, MCM enforces \textit{hierarchical feature disentanglement}, compelling the encoder to capture invariant physical-layer structures (e.g., multipath delay profiles, spatial-spectral correlations) while discarding transient noise artifacts. By training on partial observations, LWM’s transformer layers develop noise-robust latent representations, where the CLS token aggregates global signal statistics to infer missing or corrupted patches. This self-supervised objective aligns embeddings with \textbf{channel semantics} rather than pixel-level fidelity, enabling joint denoising and feature extraction without explicit noise modeling. The result is a unified framework that suppresses interference (e.g., fading, estimation errors) while preserving discriminative patterns critical for downstream tasks, bridging robustness and adaptability in dynamic wireless environments.

\subsection{LWM vs. Autoencoders Embeddings}  

Autoencoders (AEs) \cite{guo2019convolutionalneuralnetworkbased} and LWM share two core conceptual similarities despite differing in implementation: reconstruction-driven training objectives and compact latent representations. Both architectures employ reconstruction losses—AEs through pixel-level signal recovery (e.g., reconstructing raw CSI matrices) and LWM via MCM, which reconstructs masked patches by inferring contextual relationships. However, their objectives diverge in focus: AEs prioritize lossless reconstruction fidelity, while LWM emphasizes \textbf{semantic feature recovery} (e.g., multipath structure, spatial correlations) to build universal channel understanding. Second, both generate compact embeddings—AEs produce low-dimensional latent vectors, and LWM isolates a single CLS token from its full embedding sequence—but their utility differs. AE embeddings remain tightly coupled to decoder-dependent reconstruction, limiting their plug-and-play adaptability, whereas LWM’s CLS token distills global channel semantics into a task-agnostic representation, enabling direct compatibility with diverse downstream tasks (classification, regression, decision-making) without architectural overhauls. Thus, while both architectures compress inputs into compact representations, LWM’s embeddings prioritize versatile feature abstraction over pixel-perfect reconstruction, making them inherently adaptable to dynamic wireless decision-making workflows.

\section{Conclusion}
In this work, we introduced LWM, a pre-trained foundation model specifically designed for wireless communication and sensing channels. LWM draws inspiration from LLMs and Vision Transformers. Using a masked channel modeling strategy within a self-supervised framework, LWM is pre-trained to predict masked portions of the input via self-attention features and simple linear layers. Once pre-trained, LWM can generate real-time embeddings for raw channels, extracting rich, complex patterns from the input and consistently outperforming raw channels in downstream tasks. The LWM model is publicly available now.

\bibliographystyle{ieeetr}

\end{document}